\documentclass[12pt]{report}
\usepackage{epsfig,multicol,cite}
\usepackage{amsmath}
\usepackage{graphicx}
\usepackage{epsfig}

\begin{document}

\makeatletter
\renewcommand\@biblabel[1]{S{#1.}}
\renewcommand\citeform[1]{S{#1}}
\makeatother

\noindent \large {\bf Supplementary Information for
cond-mat/0610721: Potok {\em et al.} ``Observation of the
two-channel Kondo effect''}

\noindent {R. M. Potok$^{1,2,*}$, I.G. Rau$^3$, Hadas Shtrikman$^4$,
Yuval Oreg$^4$ and D. Goldhaber-Gordon$^1$}

\noindent {1 Department of Physics, Stanford University, Stanford,
California 94305.}

\noindent {2 Department of Physics, Harvard University, Cambridge,
MA, USA.}

\noindent {3 Department of Applied Physics, Stanford University,
Stanford, California 94305.}

\noindent {4 Department of Condensed Matter Physics, Weizmann
Institute of Science, Rehovot, Israel.}

\noindent {* Present Address: Advanced Micro Devices, Austin, TX.}

\ \newline \noindent {\bf Measurement techniques}

\normalsize

Measurements of differential conductance ($g = dI/dV_{ds}$) were
performed in an Oxford TLM dilution refrigerator.  (The sample is
located inside the mixing chamber.)  We measured $g$ using standard
ac lockin techniques (using PAR 124a with 116 preamp) at 337 Hz with
a RMS excitation ($V_{ex}$) of either 1$\mu V$ or 2$\mu V$,
depending on temperature ($eV_{ex} \le kT$), and measured current
with a DL Instruments 1211 preamplifier. To probe nonequilibrium
properties, we also added a dc voltage bias $V_{ds}$ to the ac
voltage $V_{ex}$ through a passive circuit. Details of the
electronics and filtering are contained in Ref.
\cite{rmpdissertation}.

\noindent \large {\bf Determination of $T_K$} \normalsize

\begin{figure}
\begin{center}
           \includegraphics[0,0][158.3mm, 61.7mm]{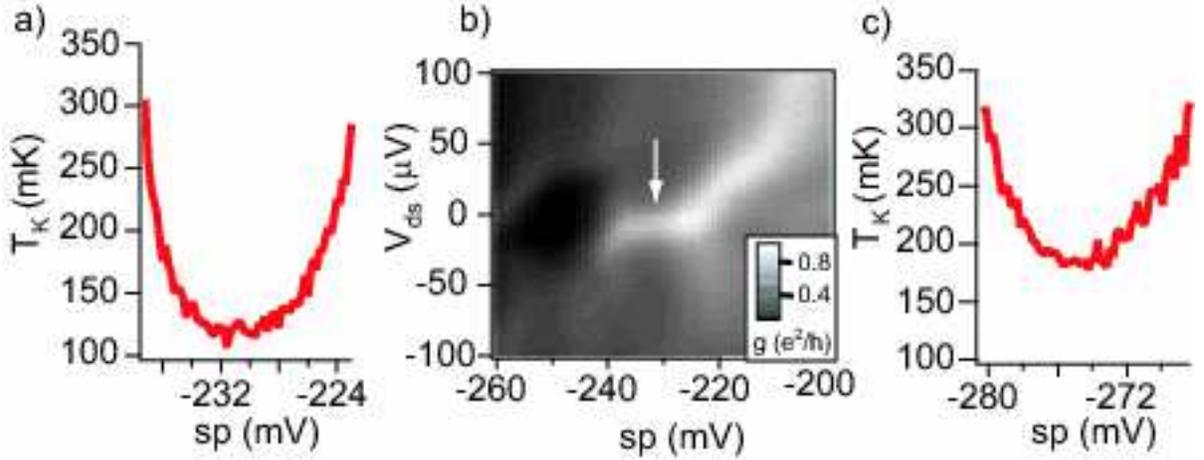}
    \caption[$T_K$ of the small quantum dot]{\label{tksupp}\footnotesize {Kondo temperature of the small quantum dot without finite
    reservoir formed.
    }}
    \end{center}
\end{figure}

When the quantum dot has an odd number of electrons and the finite
reservoir is not formed (e.g. Fig.~2 in Text), transport through the
quantum dot displays the usual signatures of Kondo effect.  The
Kondo temperature is extracted by fitting the temperature dependence
of the conductance (e.g. Fig.~2(b) inset) to Eq.~(3) in Text,
\begin{equation} \label{kt2ck}
g(T) = g_0 f(T/T_K) + a,
\end{equation}
where $f(T/T_K)$ is the expected empirical form given by
~\cite{gg.kondo.prl}.  In Fig.~\ref{tksupp}(a), $T_K$ as a function
of $sp$ is given for $c = -282\,mV$ (corresponding to a single Kondo
valley in Fig.~2(b) of Text). By measuring the conductance as a
function of $sp$ and bias voltage $V_{ds}$ in Fig.~\ref{tksupp}(b),
we observe the Kondo-enhanced density of states at the Fermi level
(marked by the arrow). In Fig.~\ref{tksupp}(c), data similar to
those in (a) are shown for a different strength of tunnel coupling
to the right lead: $c = -244\,mV$ instead of $-282\,mV$.

We do not have a detailed physical picture of the
temperature-independent constant offset $a$ in Eq.~(\ref{kt2ck}).
When left as a free fitting parameter, we find it does not vary much
across a single valley, so we choose to hold it constant as $sp$ is
varied over a single valley. At weak coupling to the right lead
(coupling gate voltage $c = -282\,$mV), we find that the offset $a =
0.21 e^2/h$ is constant over the Kondo valley near $sp=-230\,$mV. At
stronger coupling to the right lead, for the same number of
electrons in the small dot ($c = -244\,mV, sp\sim-275\,$mV) we again
find a constant offset $a = 0.09e^2/h$. In Figure~3 of the Text we
use the same two values of the offset determined before formation of
the finite reservoir, and they work fine for collapsing the data
with the finite reservoir formed. Meanwhile, $T_K$ varies
substantially across a Kondo valley (Fig.~\ref{tksupp}(a) and (c)),
reaching a minimum in mid-valley as expected and as observed in
previous
experiments~\cite{haldane,gg.kondo.prl,vanderWielWG:TheKeu}. In
addition, we note that the main conclusions of the paper are drawn
from the scaling curves from Figures 4 and 5 of the Text, which do
not depend on the values of $g_0, a$ and $T_K$, but only on the
variation in $g$ as a function of bias.

\noindent \large {\bf Conductance of the double dot system}
\normalsize

\begin{figure}
\begin{center}
           \includegraphics[width=6in]{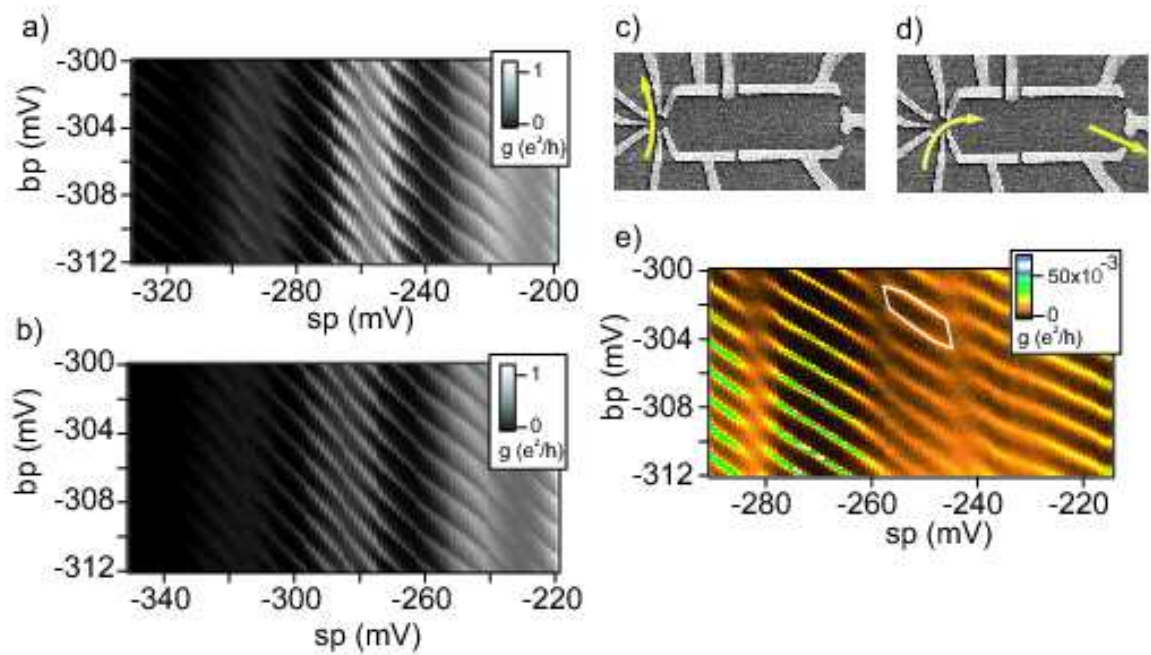}
    \caption[Conductance of double dot system]{\label{supp1}\footnotesize {Determining the charge state of the two dots.
    }}
    \end{center}
\end{figure}

In Fig.~\ref{supp1}(a),
    conductance as a function of $sp$ and $bp$ is measured through the small
    dot, along the current path shown in Fig.~\ref{supp1}(c). The main, broad conductance features
    of the small dot depend only on $sp$, as $bp$ is $> 1 \mu m$ away and thus has a very small
    capacitance to the small dot. The gate voltage $c$ is set so that the coupling
    to the finite reservoir is relatively weak. In this regime, the conductance in
    the Kondo valleys of the small
    dot, at around $sp$= -260mV and -285mV, is enhanced at low temperature.
    Gates $sp$ and $bp$ both strongly
    capacitively couple to the energy of the large dot, affecting its
    occupancy. The diagonal stripes in the conductance of the small dot
    are associated with the charge degeneracy points of the large
    dot.  Due to the large capacitive coupling between the two dots, adding an
    electron to the large dot discretely changes the electrostatic
    environment of the small dot, which changes its conductance~\cite{JohnsonAC:CouFro}.
    More complex phenomena, including SU(4) Kondo \cite{LeHurK.:Maxoas} or two
    channel Kondo physics \cite{LebanonE.:EnhtKe}, may also affect the conductance near the
    charge degeneracy points. We observe very weak temperature
    dependence in these regimes -- consistent with the exotic Kondo scenarios, but
    insufficiently distinctive to clarify the relevant physics.

    In Fig.~\ref{supp1}(b), the same type of data as in (a) is
    shown for stronger coupling between the two dots, leading
    to suppressed low-temperature conductance in the Kondo valley at
    around $sp$ = -280mV.  This same effect has also been achieved by decreasing the coupling to either of the
    two conventional leads, instead of increasing the coupling to the finite reservoir (data not
    shown.) In Fig.~\ref{supp1}(e) conductance as a function of $sp$ and $bp$
is measured through both
    dots in series, along the current path shown in Fig.~\ref{supp1}(d). The charge degeneracy
    points for both the large and small quantum dots are apparent from
    these data, revealing the charge stability hexagon (white
    hexagon superimposed as a guide to the eye).

\newpage \noindent \large {\bf Impact of applied magnetic field}
\normalsize

As noted in the Text, we applied a magnetic field normal to the
plane of the sample in order to deflect electron trajectories so
that they cannot travel directly between the entry and the exit
point contacts to the small quantum dot. Such direct paths,
coexisting with resonant tunneling, give rise to Fano resonances, as
seen in previous experiments on quantum dots~\cite{joern.fano}, and
in measurements on the present sample at zero magnetic field.
Manipulating wavefunctions with modest normal magnetic fields has
been used in many other realizations of Kondo effect in quantum
dots, for example the achievement of the unitary limit of transport
in~\cite{vanderWielWG:TheKeu} and tuning of Kondo coupling in a
3-terminal ring~\cite{LeturcqR:ProKds}.

How much impact should this field have on the Kondo physics? Due to
the small g factor in GaAs dots, the Zeeman splitting is quite small
at the magnetic field we applied (130 mT). Theoretically, magnetic
field should be a relevant perturbation to the 2-channel Kondo state
(See for example the discussion after Eq.~(47) in
\cite{glazman2ck}). However, this should only be the case at very
low temperatures $kT<E_{\rm Zeeman}^2/kT_K$. The field applied in
the experiment $B=0.13 \;Tesla$ yields a Zeeman energy $E_{\rm
Zeeman}= 0.13 * 25 =3.25 \mu e V$ for GaAs. For $kT_K \approx 12 \mu
{\rm eV}$ in our experiment, the temperature below which the
magnetic field should be relevant is $E_{\rm Zeeman}^2/kT_K \approx
.8 \mu {\rm eV} \approx 9.5$ mK. This effective energy scale is
comparable to the base temperature of the experiments that was
around $12mK$, so for 2CK we are never in the regime where magnetic
field is a relevant perturbation, with the possible exception of our
very lowest temperature. The magnetic field is probably even less
important than this calculation suggests, since $E_{\rm Zeeman}$ is
often lower by a factor between 1.3 and 3 in a GaAs/AlGaAs
heterostructure, where the electron wavefunction leaks into the
AlGaAs barrier.

In future, it would be interesting to observe the effect of a larger
magnetic field, which should perturb the 2CK state. We would want to
apply the field in the plane of the 2DEG, to minimize its effects on
orbital states. Applying a field precisely in-plane is non-trivial.
We are now setting up to do those measurements.

Note: For 1CK, which is not the main focus of the present work,
Zeeman coupling is not a relevant perturbation in the renormalizatin
group sense. Provided Zeeman energy is substantially smaller than
Kondo temperature (as it is in our experiments) it should have an
effect similar to that of bias or temperature $(G = G_0 - {\rm
const}_1 \cdot(T/T_K)^2) - {\rm const}_2 \cdot(eV_{\rm ds}/kT_K)^2)
- {\rm const}_3 \cdot(g\mu_B B/k T_K)^2)$, where the three constants
are all of order unity. In our experiments, $g \mu_B B$ is roughly
one to three times $kT_{\rm base}$, depending on the exact g-factor.
Since all the perturbations are substantially smaller than the Kondo
temperature, the presence of the magnetic field should not
substantially affect the scaling of conductance with temperature and
bias.

\noindent \large {\bf Scaling analysis of 1CK data} \normalsize

\begin{figure}
\begin{center}
           \includegraphics[width=6in]{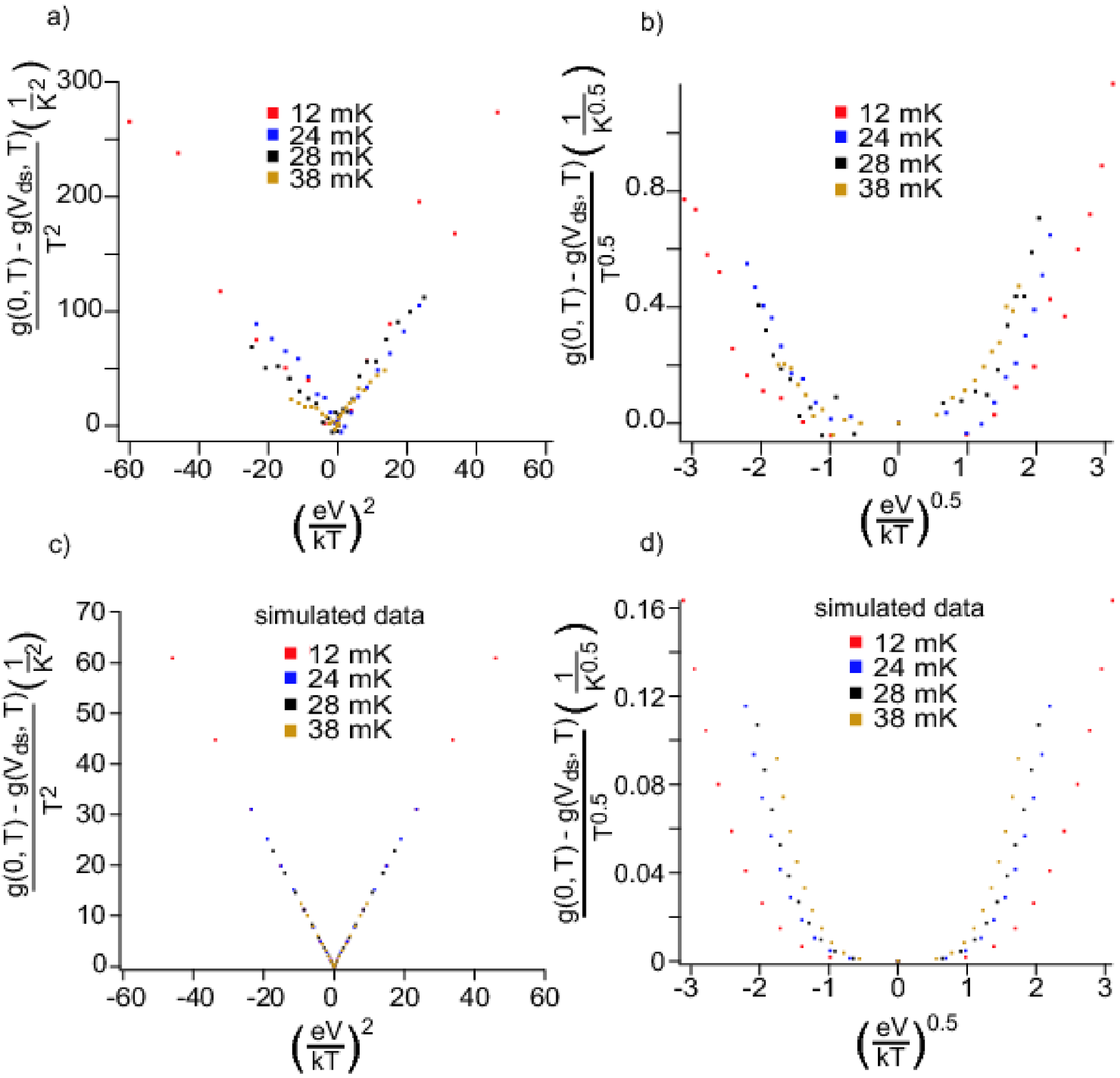}
    \caption[Scaling of 1CK]{\label{supp2}{\footnotesize Scaling of
    1CK
    }}
    \end{center}
\end{figure}

In the main Text we demonstrated that the data we identify as
reflecting a symmetric 2CK state cannot be described by a Fermi
liquid scaling appropriate to 1CK. It is important to establish the
converse: that the data we identify as reflecting 1CK do not follow
2CK scaling. We show this in Fig.~\ref{supp2}, where the 1CK data
presented in Figure 4 of the Text are seen to scale as expected for
1CK and not as expected for 2CK. From Eq.~(4) of the Text, the
expected scaling for 1CK is
    \begin{equation} \label{sc1ck2ck}
\frac{g(0, T) - g(V_{ds}, T)}{T^\alpha} = \kappa \left(
\frac{eV_{ds}}{kT} \right) ^2,
\end{equation}
with $\alpha = 2$ and $\kappa = 0.82
\frac{g_0}{T_K^2}$~\cite{TDelta}. In Fig.~\ref{supp2}(a), we show
the same 1CK scaling plot as in Fig.~4(d) of the Text, but without
normalizing to account for $T_K$ or $g_0$.

    In Fig.~\ref{supp2}(b), we scale the same data from Fig.~\ref{supp2}(a)
    as would be appropriate for 2CK behavior, i.e. with $\alpha = 0.5$.
    In Fig.~\ref{supp2}(c) and (d) we simulate idealized 1CK (Fermi liquid) data
    and scale them as would be appropriate for 1CK ($\alpha = 2$, (c)) and 2CK ($\alpha=0.5$, (d).
    Comparing Fig.~\ref{supp2}(b) and (d), the simulated 1CK data deviate from perfect
    2CK scaling very similarly to how the actual 1CK data deviate. Note that the
    qualitative behavior is the opposite of what one would expect from a breakdown of scaling
    when approaching some finite energy scale (e.g. $T_K$): curves at higher temperatures
    fall inside those at lower temperatures, instead of ``peeling off'' toward the outside
    above a certain bias voltage. The nonlinear fits presented in the Text quantify these
    observations: the best fit for $\alpha$ is $1.72 \pm 0.4$ for the 1CK data and $0.62 \pm 0.21$ for the symmetric
    2CK data, clearly distinguishable from each other, and both consistent with theoretical
    expectations ($\alpha = 2$ and $0.5$, respectively.)

\noindent \large {\bf Comparison of theory and experiment for
$\kappa$, 1CK scaling prefactor} \normalsize

As noted in the Text, the value of $\kappa$ depends on the
underlying model (Kondo effect can be derived from various different
models), numerical calculations ($\kappa$ connects low-energy
behavior to high-energy behavior, and no analytical results can make
this link quantitatively), and proximity to the symmetric 2CK point
(near the symmetric point, $T_{\rm K}$ is replaced by $T_{\Delta}$,
a measure of the asymmetry). Here we outline how to determine
$\kappa$ theoretically, and we comment on the link to our
experimental result.

First, a Kondo energy scale (or Kondo temperature) is only a
crossover scale, so different definitions could yield values
differing by some constant multiple. We want results that are
independent of these initial definitions. Theoretically, the Kondo
temperature is usually defined in terms of a thermodynamic quantity
such as susceptibility rather than a dynamic quantity such as
electrical conductance, so we must use a model to link the two.
According to Costi~\cite{costiprivatecomm},

$$g(T)=g_0\left(1-
                  \frac{\pi^4}{16}
                  \left(\frac{T}{T_0}\right)^2\right),$$~\cite{conductanceoffset}

where $T_0$ is defined according to

            $$\chi(T=0) =  \frac{(g \mu)^2 }{4 k T_0}.$$

Now
$$\kappa = g_0 \left(\frac{\pi^2 }{4 T_0}\right)^\alpha  \frac{3}{2 \pi^2}$$
For the case $\alpha=2$ we find
$$\kappa = g_0 \frac{3 \pi ^4 }{32\pi^2 T_0^2} = \frac{3 \pi^2}{32} \frac{1}{T_0^2}. $$

Next, we must link the thermodynamically-defined Kondo scale $T_0$
to $T_{\rm K}$, defined according to $g(T_{\rm K}) =
g_0/2$~\cite{conductanceoffset}. Costi's NRG calculations suggest
that this link depends mildly on details of the system such as the
dimensionality of the leads. For 2D leads, $T_{\rm K} = 0.94 T_0$.
This yields $\kappa = 0.82 \frac{g_0}{T_K^2},$ as reported in the
Text. This is in rough but satisfactory agreement with our
experimentally-extracted value $\kappa = 0.25$ for both 1CK with the
conventional leads and 1CK with the finite reservoir. Note that
other approaches to the basic Kondo model may or may not give the
same result. A. Schiller's calculations based on an exactly-solvable
model at the Toulouse limit give
$$g(T)=g_0\left(1-
                  \frac{\pi^4}{48}
                  \left(\frac{T}{T_0}\right)^2\right),$$
yielding a value of $\kappa$ three times smaller than that of the
other models, and in almost perfect agreement with our experimental
results. Apart from this (perhaps serendipitous) match we have no
reason to believe that the exactly-solvable model at the Toulouse
limit is a better description of the low-energy properties of our
system than Nozi\`{e}res's Fermi liquid approach.

A final complication in quantitative comparison of theory and
experiment is that our measurements are not very far from the
symmetric 2CK, so $T_{\rm K}$ should be replaced by $T_{\Delta}$.
It's not clear whether $T_\Delta$ should act the same as $T_{\rm K}$
at {\em both} low and high energies. Therefore, it will be
interesting to perform these same analyses on a two-lead dot which
exhibits simple 1CK behavior, with no link to 2CK.

\noindent \large {\bf Raw data for 2CK scaling analysis} \normalsize

\begin{figure}
\begin{center}
           \includegraphics[0,0][60mm, 62mm]{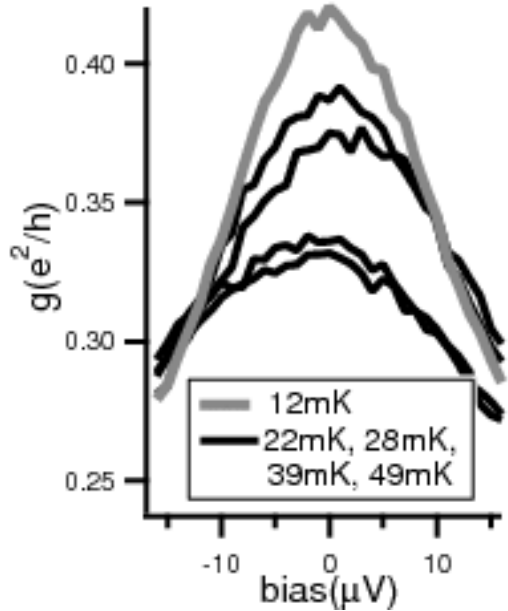}
    \caption[Unscaled 2CK]{\label{unscaled}\footnotesize {Unscaled
    conductance curves at different temperatures at the 2CK point.
    }}
    \end{center}
\end{figure}

Fig.~\ref{unscaled} shows the raw data used in the 2CK scaling
analysis. These data were obtained under conditions similar to those
for the $c=-260\,$mV curve in Fig.~5(e) in the Text, which shows
differential conductance at widely-spaced values of the coupling
gate voltage $c$. Since the parameters of the system had shifted
since acquisition of the data in Fig.~5(e), the coupling gate had to
be changed to $c=-258\,$mV. For the scaling analysis (Fig.~5(f) of
Text), to reduce the noise in the value of $g(V_{\rm ds}=0)$ we
averaged the conductances at $-1 \mu$V, $0$, and $1 \mu$V.

\noindent \large {\bf Match of raw data to 2CK predictions}
\normalsize

\begin{figure}
\begin{center}
           \includegraphics[0,0][72mm, 82mm]{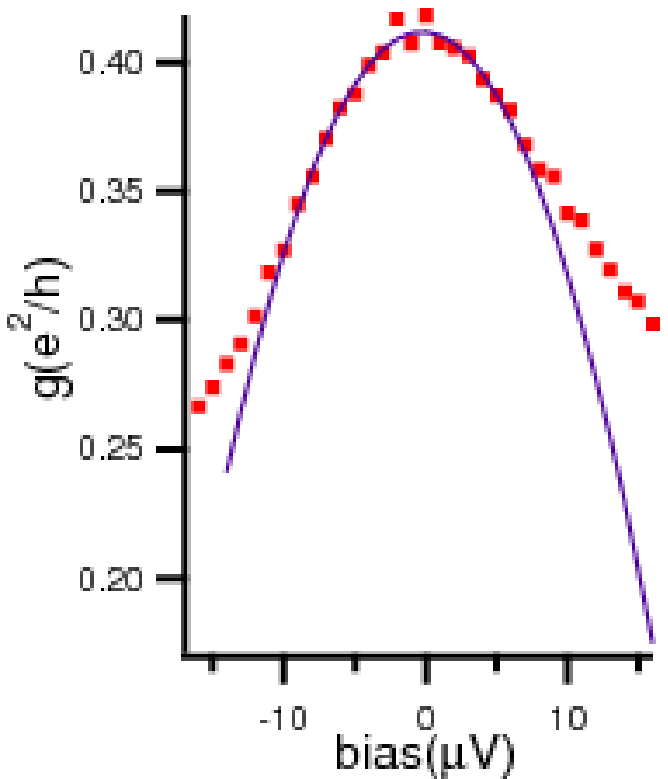}
    \caption[2CKparabola]{\label{parabola}\footnotesize {Parabolic fit to $g(V_{\rm ds})$ at small bias.}}
    \end{center}
\end{figure}

\begin{figure}
\begin{center}
           \includegraphics[0,0][72mm, 82mm]{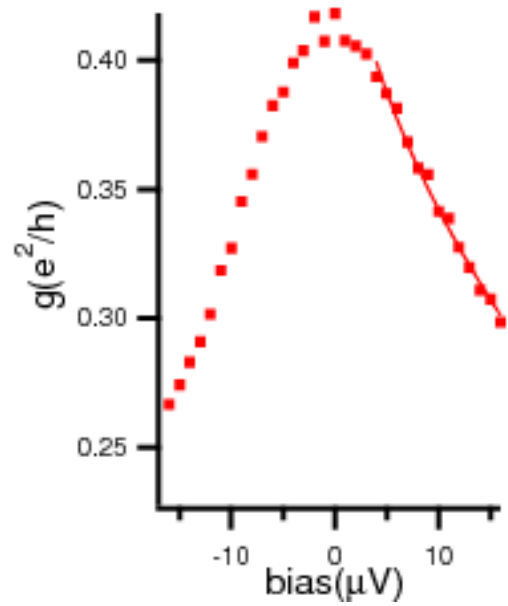}
    \caption[2CKsquareroot]{\label{squareroot}\footnotesize
    {Square-root fit to $g(V_{\rm ds})$ at intermediate bias.}}
    \end{center}
\end{figure}

Fig.~\ref{parabola} shows the 12 mK raw data used in the 2CK scaling
analysis (gray curve from Fig.~\ref{unscaled}.) A parabolic fit
works only at low bias. In contrast, a square-root fit ($g(V_{\rm
ds})=g_0 - g_1 \sqrt{V_{\rm ds}}$, with $g_0$ and $g_1$ as fit
parameters, Fig.~\ref{squareroot}) works well at intermediate bias
($V_{\rm ds} = 5$ to $15 \mu$V) This crossover from quadratic to
square-root behavior at bias a few times $kT$ agrees with conformal
field theory predictions for
2CK~\cite{AffleckI.:Exacrm,glazman2ck,vonDelftJ:The2Km}. This match
is reinforced by the more complete scaling analysis in Fig.~5(f).

\noindent \large {\bf Asymmetry of coupling to the two conventional
leads that comprise the ``infinite reservoir''} \normalsize

The tunnel barriers between the local site and the two conventional
leads were intentionally tuned to be asymmetric, for two reasons:

a. Existing theoretical calculations for 2CK (and 1CK) give density
of states {\em in equilibrium}. Our scaling measurements involve
applying finite bias from one lead to the other. The strong
asymmetry between the couplings to the two leads means that the
local site remains essentially in equilibrium with the more
strongly-coupled lead, validating quantitative comparison with
predictions.

b. The symmetric 2CK state occurs when $J_1=J_2$, which requires
that $\Gamma_{\rm fr} \approx \Gamma_{\rm ir}$, where $\Gamma_{\rm
ir}$ is the sum of the tunnel rates to the two conventional leads.
%In fact, Coulomb charging in the finite reservoir means that the
%tunneling to the finite reservoir must be even stronger: near the
%center of a charging hexagon the symmetric state requires
%$\Gamma_{\rm fr} \approx 1.3 \Gamma_{\rm ir}$.
If all three tunnel barriers were equal, we would instead have
$\Gamma_{\rm fr} = 0.5 \Gamma_{\rm ir}$. It turns out to be
easiest to tune the system by first matching the tunnel barrier of
the finite reservoir to that of one of the open leads. With the
finite reservoir open to the outside world (gate $n$ grounded) and
one conventional lead fully pinched off, we maximized the
two-terminal conductance between the other conventional lead and
the infinite reservoir (in fact we found it could be very near
$2e^2/h$, usually $\sim 1.8 e^2/h$). This means $\Gamma_{\rm fr}
\approx \Gamma_{\rm ir}$. We then slowly cracked open the second
conventional lead and closed off the finite reservoir from the
outside world (using gate $n$). In this way, we maintained
$\Gamma_{\rm fr}$ near $\Gamma_{\rm ir}$, as needed for $J_1 =
J_2$. We felt this was the best method to ensure nearly equal
$\Gamma$s and $J$s.

\bibliography{main2006c}

\begin{thebibliography}{10}
\bibitem{TDelta}
This equation describes the low-energy \protect{F}ermi-liquid behavior, at an
  energy scale substantially below \protect{$kT_{\rm
  K}$}~\cite{MajumdarK:NonKiP}. for a channel-asymmetric \protect{2CK} system,
  the crossover scale \protect{$T_{\rm K}$} is replaced by
  \protect{$\min\{T_{\rm K},T_\Delta\}$}, where \protect{$T_\Delta$} is a
  measure of the channel asymmetry which goes to zero when
  \protect{$J_1=J_2$}~\cite{glazman2ck}.
\bibitem{conductanceoffset}
Because we have a temperature-independent conductance offset, we replace
  \protect{$g(T)$} with \protect{$\tilde{g}(T) \equiv g(T) - a$} in this
  formula, for comparison with the data on \protect{1CK} with the conventional
  leads (fig.~4(d) of text) -- the analysis is similar for \protect{1CK} with
  the finite reservoir, but the temperature-dependence has the opposite slope.
\end{thebibliography}
\bibliographystyle{unsrt}

\end{document}